\documentclass[aps,prb,twocolumn,groupedaddress]{revtex4-1}

\usepackage{graphicx}

\begin{document}

\title{Comment on "Self-doping effects in cobalt silicide CoSi: Electrical, magnetic, elastic,
and thermodynamic properties"}

\author{V. N. Narozhnyi}
\email{narozhnyivn@gmail.com}
\affiliation{Institute for High Pressure Physics, Russian Academy of Sciences, 142190, Troitsk, Russia}

\date{\today}

\begin{abstract}
In this Comment it is argued that Stishov \emph{et al.} [Phys.\ Rev.\ B \textbf{86}, 064433 (2012)] incorrectly estimated concentrations of (supposed) paramagnetic centers with $\mu_{\text{eff}}\approx4.8\,\mu_{\text{B}}$ in the investigated CoSi crystals. Correct estimation gives concentrations of such centers from 25 to 50 times smaller than reported ($\sim0.04\div0.16\,\%$ instead of $\sim2\div4\,\%$). Also the reported data on temperature dependence of resistivity $\rho(T)$ of four CoSi crystals prepared in different Labs are so close to each other at $T\approx250\div300\,\text{K}$ that it is extremely unlikely to be reproducible for any reasonable accuracy of resistivity measurements. These and some other problems of the paper are related to the key points of the authors argumentation. As a result their main conclusions become unjustified.
\end{abstract}


\maketitle

In a recently published paper, Stishov \emph{et al.}\cite{PhysRevB.86.064433} have presented results on electrical, magnetic, elastic, and thermodynamic properties of CoSi single crystals. In this Comment it is shown that there are significant problems with the reported data as well as with the proposed interpretations at least for electrical and magnetic properties.

Concerning \textit{magnetic properties} of CoSi, Stishov \textit{et al.}\cite{PhysRevB.86.064433} have reported some data on temperature dependence of magnetic susceptibility $\chi (T)$ for four single crystals prepared in different Labs. To analyze the $\chi(T)$ curves having clear minima for all samples the authors used the expression $\chi (T) = \chi_0 + \text{D} \times T + \text{C}/(T - \Theta)$, where the first two terms are supposed to be connected with a diamagnetic contribution, whereas the third term is a Curie-Weiss contribution.\footnote{Actually this expression in Ref.~\onlinecite{PhysRevB.86.064433} is mistakenly written as $\chi (T) = \chi_0 + \text{D} \times T + \text{C}/(T + \Theta)$} Although this expression gives a possibility to fit the experimental data rather well, it should be underlined that the second term in it was introduced without any physical justification (\textit{ad hoc}).

\begin{figure}[b]
\includegraphics{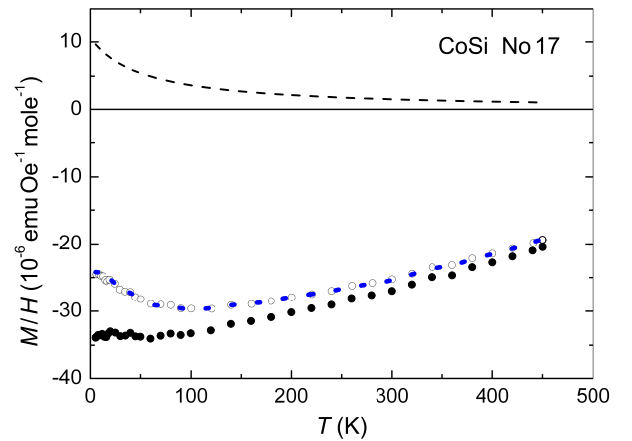}%
\caption{\label{Fig1} (Color online) (1) - $M/H$ vs.~{T} for CoSi crystal No~17 in magnetic field $H=10~\text{kOe}$  (open symbols). The data are the same as shown in Fig.~4 of Ref.~\onlinecite{PhysRevB.86.064433} for the sample marked as Br17. (2) - Paramagnetic contribution to $M/H$ (dashed line in the upper part of a graph). (3) - $M/H$ of a \textit{hypothetical} "ideal" CoSi sample (full symbols). (4) - The sum of (2) and (3) shown as a dotted line going through open symbols.}
\end{figure}

The same experimental data \footnote{The reported in Ref.~\onlinecite{PhysRevB.86.064433} results are connected with magnetization $M$ linearly dependent on magnetic field $H$. Actually raw experimental data contain some nonlinear terms in $M(H)$ which were subtracted after careful $M(H)$ measurements\cite{Our_Conf,Our_Paper} at various $T$.} have been analyzed using an approach without any \textit{ad hoc} assumptions\cite{Our_Conf,Our_Paper}. This analysis (based on a comparison of the data for at least two samples with considerably different Curie-Weiss contributions to $\chi (T)$, see Refs.\,\onlinecite{Our_Conf,Our_Paper} for details) gives a possibility to extract a magnetic susceptibility of a \textit{hypothetical} "ideal" CoSi crystal (containing no paramagnetic centers, defects, etc.). Some of these results are shown in Fig.~\ref{Fig1}. Magnetic susceptibility of a hypothetical "ideal" CoSi (shown by full circles) is diamagnetic at $T=5.5\div450\,\text{K}$. At high temperatures $\chi (T)$ dependence of an "ideal" CoSi is close to linear, but at low $T$ it flattens. A character of $\chi (T)$ of an "ideal" CoSi is not very sensitive to a selection of samples for such an analysis, therefore diamagnetic $\chi (T)$ dependence shown on the Fig.~\ref{Fig1} for a hypothetical "ideal" CoSi can be considered as \textit{intrinsic} for CoSi,\cite{Our_Conf,Our_Paper} contrary to the conclusion of Ref.~\onlinecite{PhysRevB.86.064433} in which $\chi (T)$ with a transition from diamagnetic to paramagnetic on cooling is considered to be intrinsic.

This analysis gives also a possibility to determine more reliably Curie-Weiss contributions to $\chi (T)$ (and, hence, the Curie constants) of "real" investigated samples. As an example, a dashed line in the Fig.~\ref{Fig1} represents a paramagnetic term for the sample No~17. The sum of $\chi(T)$ dependence of an "ideal" CoSi and a paramagnetic Curie-Weiss term excellently fits the experimental data.

It should be also mentioned here that an idea of a "generation" of magnetic moments in CoSi on cooling\cite{PhysRevB.86.064433} is not consistent with an excellent approximation of a paramagnetic contribution to $\chi (T)$ given by the Curie-Weiss formula. Naturally, in a case of "generation" of magnetic moments, i.e., when magnetic moments strongly depend on temperature, a $\chi (T)$ should considerably deviate from a dependence given by the Curie-Weiss expression.

Although the values of the Curie constants determined using this approach\cite{Our_Conf,Our_Paper} [$\text{C}=(3.7;\, 4.9;\, 2.8;\, 0.84)\times10^{-3}\, \text{(emu\,K/mole\,Oe)}$ for the samples Ames, Ural, Br144, Br17, respectively (in notation of Ref.~\onlinecite{PhysRevB.86.064433})] are to some extent different from the reported by Stishov \textit{et al.},\cite{PhysRevB.86.064433} \textit{the main problem} related with magnetic properties reported in Ref.~\onlinecite{PhysRevB.86.064433} is not connected with this moderate difference but with an \textit{incorrect method} of estimation of concentrations of (supposed) $\text{Co}^{2+}$ paramagnetic centers (with an effective magnetic moment $\mu_{\text{eff}}$ of about $4.8\,\mu_{\text{B}}$).

Direct calculation of concentrations of such centers from the reported by Stishov \textit{et al.}\cite{PhysRevB.86.064433} Curie constants [$(2\div8)\times10^{-3}\,\text{(emu\,K/mole\,Oe)}$, mole in Ref.~\onlinecite{PhysRevB.86.064433} is missed] using an expression for $\text{C}$ for a diluted magnetic system $\text{C}=x\text{N}_{\text{A}}\mu_{\text{eff}}^2/3\,\text{k}_{\text{B}}$ ($\text{N}_{\text{A}}$ is Avogadro's number, $\text{k}_{\text{B}}$ - Boltzmann's constant and $x$ is a concentration of paramagnetic centers, see, e.g., Ref.~\onlinecite{Selwood_book_1956}), gives values $\sim0.04\div0.16\,\%$. These are from 25 to 50 times smaller than obtained by Stishov \textit{et al.}\cite{PhysRevB.86.064433} ($\sim2\div4\,\%$) using their "two steps" method. Even much smaller mistake in estimation of magnetic properties (as, e.g., a mistake in 1.4 times in determination of $\mu_{\text{eff}}$) can in some cases completely demolish arguments of an original interpretation\cite{Narozhnyi_1999}.

It is easy to see that a miscalculation in Ref.~\onlinecite{PhysRevB.86.064433} is connected with \textit{linear} (instead of \textit{quadratic}) scaling when concentrations were estimated from the effective numbers of magneton per formula unit.

It is clear that a rather small concentration (namely $\sim0.04\div0.16\,\%$) of supposed paramagnetic centers with $\mu_{\text{eff}}\approx4.8\,\mu_{\text{B}}$ is sufficient to explain the observed Curie-Weiss contributions to $\chi(T)$ of the CoSi samples investigated in Ref.~\onlinecite{PhysRevB.86.064433}. Naturally, before any discussions of "self-doping effects"\cite{PhysRevB.86.064433}, a simplest possible explanation (connected with a presence of magnetic impurity of some kind, e.g., $\text{Fe}^{3+}$ impurity\footnote{A problem of magnetic moment formation on Fe impurity in $\text{Co}_{1-x}\text{Fe}_x\text{Si}$ is rather complicated, see, e.g.,   Refs.~\onlinecite{Kawarazaki_1976,Beille_1983,Guevara_2004}. It is not possible to discuss it in more detail here. Our estimation of supposed iron concentration with $\mu_{\text{eff}}\approx5.9\,\mu_{\text{B}}$ from the values of Curie constants determined in Ref.~\onlinecite{Our_Paper} gives the values $\sim0.019\,\%$ for the sample No\,17 and $\sim0.064\div0.11\,\%$ for the other samples.} with  $\mu_{\text{eff}}\approx5.9\,\mu_{\text{B}}$) should be excluded. This was easy to ensure for relatively large concentrations $\sim2-4\,\%$ miscalculated in Ref.~\onlinecite{PhysRevB.86.064433}, but it become substantially more difficult for considerably smaller impurity content obtained in an analysis described above.

An estimation of an actual impurity concentration in the investigated samples performed by arc atomic emission spectroscopy (AES) have shown\cite{Our_Paper}, e.g., that concentration of iron in the CoSi sample No\,17 is $\sim(0.02\pm0.01)$~mass\,\%. Therefore it is not excluded that Fe impurity can be solely responsible for a paramagnetic contribution to $\chi (T)$ of this particular sample\cite{Our_Paper}. Iron impurity in the Br17 crystal can also give a natural explanation (connected with Kondo effect) for a shallow minimum in $\rho(T)$ as well as for a small negative magnetoresistance reported for it in Ref.~\onlinecite{PhysRevB.86.064433}.

It should be also noted that the Curie-Weiss behavior in $\chi(T)$ of CoSi does not necessarily imply an existence of \textit{local} magnetic moments. It is sufficient to mention MnSi\,\cite{Moriya_book_1985} and closely related $\text{Co}_{1-x}\text{Fe}_x\text{Si}$ alloys\cite{Kawarazaki_1976,Beille_1983,Guevara_2004}. These are compounds with strong paramagnetic $\chi (T)$ dependencies, which are usually considered as connected with spin-fluctuations of band electrons.

The next problem of Ref.~\onlinecite{PhysRevB.86.064433} is connected with \textit{temperature dependencies of resistivity} $\rho(T)$ of four different CoSi crystals. The reported data are so close to each other (within  $\sim1\,\%$) at $T\approx250\div300\,\text{K}$ that it is extremely unlikely to be reproducible for any reasonable accuracy of resistivity measurements.

To demonstrate this it should be mentioned that usually an uncertainty in $\rho$ measurement is mainly connected with an uncertainty of a geometrical factor. This is especially true for $\rho$ measurements of relatively small single crystals as well as for experiments under pressure. For real samples with relatively small sizes a typical accuracy in determination of $\rho$ may be considered as $\sim20\,\%$. To achieve a better result a very careful measurements of a sample dimensions are necessary. Also it is essential to take into account finite dimensions of electrical contacts as well as possible nonhomogeneous character of current flow through the sample, etc. Taking all these points into consideration it is very difficult to understand the very close values of reported resistivity for the four different CoSi crystals at $T\approx250\div300\,\text{K}$. Moreover, careful examination of Fig.\,1 from Ref.~\onlinecite{PhysRevB.86.064433} has shown that $\rho(T)$ curves coincide within $\sim0.25\,\%$ for all crystals near $T=273\,\text{K}$. A rough estimation (given by elemental statistical analysis) of a probability of such coincidence shows that it is extremely low, namely $\sim2\times10^{-7}$, \textit{even in case of ideally equal resistivity of all four samples}. (A probability remains very small, $\sim2\times10^{-5}$, in case of much better accuracy of resistivity measurements, $\sim5\,\%$\,, which is really hard to achieve for real crystals.)

It is natural to ask, whether it is possible to reproduce the reported results on $\rho(T)$? An answer is very simple: a probability to get similar results in two subsequent \textit{independent} $\rho$ measurements of four samples (that means making new contacts, etc.) is of an order of  $(2\times10^{-7})^2\approx4\times10^{-14}\,$ ($\approx4\times10^{-10}\,$ for $\sim5\,\%$\, accuracy of resistivity measurements). Physically this event can be considered as almost impossible, i.e., \textit{a reproducing of a surprising coincidence in resistivity reported in Ref.~\onlinecite{PhysRevB.86.064433} is practically impossible}. In the best case it should be considered as an accidental event. [Actually some other reasons, e.g., a normalization of the reported $\rho(T)$ curves at $T\approx273\,\text{K}$ for some reason unmentioned in the paper, are far more probable than an accidental coincidence.]

A nice illustration for the above discussion can be obtained by comparison of Figs.\,1 and 2 from Ref.~\onlinecite{PhysRevB.86.064433}. The Fig.\,2 represents $\rho(T)$ curves for Ames CoSi crystal determined at various pressures, including results at normal pressure. Data for \textit{the same} CoSi crystal are also shown on the Fig.\,1. It is easy to see a difference in resistivity of \textit{two samples of the same crystal} approaching $\sim15\,\%$ and $\sim50\,\%$ at $T=300$ and $5\,\text{K}$ respectively.

It is unreasonable to discuss any questions connected with a comparison of $\rho$ for \textit{different} crystals as well as a problem of an applicability of the parallel resistor model (based on "practically the same" high temperature asymptotic values of resistivity \cite{PhysRevB.86.064433} of different samples of CoSi), etc., when the data for \textit{two samples from the same crystal} vary from $\sim15\,\%$ up to $\sim50\,\%$.

In a conclusion, the problems of Ref.~\onlinecite{PhysRevB.86.064433} discussed above concern key points of the authors argumentation. As a result main conclusions of this paper become unjustified.

\begin{acknowledgments}
Valuable discussions with V. N. Krasnorussky are greatly appreciated.
\end{acknowledgments}

\bibliography{CoSi_Comment_v4}

\end{document}